
%
%

\input PHYZZX.TEX

\font\largemath = cmti10 scaled \magstep2
\textfont4=\largemath
\mathchardef\lS="0453
\footline={\hfill}

\pubnum={CHIBA-EP-64}
\date={October 12, 1992 revised}
\titlepage
\title{ Dynamical Symmetry Breaking on Langevin Equation :\break
Nambu $\cdot$ Jona-Lasinio Model }

\author{Kenji Ikegami}

\address{Graduate School of Science and Technology,  \break
                 Chiba University, \break
        1-33 ~Yayoi-cho,  ~Inage-ku, ~Chiba 263, JAPAN }%

\andauthor{Riuji Mochizuki and Kazuhiro Yoshida}

\address{ Department of Physics, Faculty of Science,\break
  ~Chiba University, \break
   1-33 ~Yayoi-cho,  ~Inage-ku, ~Chiba 263, JAPAN }%

\baselineskip = 12pt

\abstract{In order to investigate dynamical symmetry breaking, we study
Nambu$\cdot$Jona-Lasinio model in the large-N limit in the stochastic
quantization method. Here in order to solve Langevin equation, we impose
specified initial conditions and construct ``effective Langevin equation''
in the large-N limit and give the same non-perturbative results as
path-integral
approach gives. Moreover we discuss stability of vacuum by means of
``effective potential''.  }%

\baselineskip = 18pt
%

\REF \parisi{G.Parisi and Y.Wu, Sci.Sin. {\bf 24} (1981)483}
\REF \damgaad{For a review, P.H.Damgaad and H.H\"uffel, Phys.Rep.
{\bf 152}(1987)227}
\REF \ghost{M.Namiki, I.Ohba, K.Okano and Y.Yamanaka, Prog.Theor.Phys.{\bf
69}\break (1983)1580 \hfill}
\REF \nakghost{A.Nakamura, Prog.Theor.Phys. {\bf 86}(1991)925}
\REF \higashi{K.Higashijima, Prog.Theor.Phys.Suppl. {\bf 104}(1991) and
reference there in}
\REF \ito{M.Ito and K.Morita, Prog.Theor.Phys. {\bf 87}(1992)207 and
reference there in}
\REF \mocyosi{R.Mochizuki and K.Yoshida, CHIBA-EP-63(1992)}
\REF \nambu{Y.Nambu and G.Jona-Lasinio, Phys.Rev.{\bf 122}(1961)345}
\REF \gross{D.J.Gross and A.Nevou, Phys.Rev.{\bf D10}(1974)3235}
\REF \sakita{B.Sakita, in Proc.Johns Hopkins Workshop 7,
eds.Domokos and Kovesi-\break Komokos (World Scientific, Singapore,
1983);\hfill \break
K.Ishikawa, Nucl.Phys.{\bf B241}(1984)589 \hfill}

\footline={\hfill-- \folio\ --\hfill}
\pagenumber = 1
\chapter{Introduction}
 The stochastic quantization method (SQM) was first proposed by Parisi and
Wu as an alternative quantization method.
\refmark\parisi \refmark\damgaad
They showed that this method can be applied to gauge theories without the
gauge fixing procedure, i.e. without Faddeev-Popov ghost fields. In spite
of introducing no ghost field, this method produces the same contribution
as the path-integral method gives. This is also confirmed perturbatively for
Yang-Mills fields
\refmark\ghost
and recently for non-Abelian anti-symmetric tensor fields.
\refmark\nakghost

Recently, dynamical symmetry breaking has been studied in SQM.\refmark\ito
Their main tool for investigation is the stochastic generating functional
from which effective potential is derived. On the contrary, the Langevin
equation, which is very useful in perturbative calculation, has lost its
position because dynamical symmetry breaking appears as a
non-perturbative phenomenon. Nevertheless if we can use it even for
non-perturbative calculation, we may expect that computations will become
easier and the relation between perturbative and non-perturbative phenomena
will become clearer. For this purpose, in the previous paper \refmark\mocyosi
we introduced the ``effective Langevin equation'' for the two-dimensional
non-linear $\sigma$ model in the large-N limit and gave the same results as
the path-integral approach gave. This effective Langevin equation,
which is constructed in the large-N limit from the ordinary Langevin equation,
is
a powerful tool for non-perturbative calculation. In this paper we introduce an
assumption and impose initial conditions in order to solve the effective
Langevin equation and apply this effective Langevin equation to
Nambu$\cdot$Jona-Lasinio model \refmark\nambu in the large-N limit and show
that
this approach gives the same results as the path-integral approach
gives.\refmark\higashi

Moreover we discuss stability of vacuum by using ``effective potential''.
This effective potential is constructed at any fictitious time and coincides
with one given in the path-integral method in the large-N limit. Of course, we
obtain the same masses as the effective potential in the path-integral method
gives.

This paper is organized as follows. In section 2, we derive
the effective Langevin equation for Nambu$\cdot$Jona-Lasinio model and in
order to show coincidence with other methods, we calculate the scalar
and pseudoscalar propagators. In section 3, we discuss stability of
vacuum. In section 4, we give summary.

\chapter{Effective Langevin Equation}
We derive ``effective Langevin equation'' in this
section. By using this equation, we show later that non-perturbative results
are obtained. At first, the classical Euclidean action for Nambu $\cdot$
Jona-Lasinio model\refmark\nambu in four dimensions is \refmark\gross
  $$
    S = \int d^4x\ \lbrack\overline{\psi}^i \lbrace i
         \gamma_{\mu}\partial_{\mu}
       + g(\sigma + i \gamma_5 \pi) \rbrace \psi^i
       + {N M^2 \over 2}(\sigma^2 + \pi^2)\rbrack,\eqno(2.1)
     $$
      $$
         i=1,2,\cdots,N,
  $$
where $\sigma$ and $\pi$ are auxiliary fields and $M$ is a mass scale which
is introduced due to make the coupling constant $g$ dimensionless. This
action is invariant under global transformation
$$
\psi \rightarrow e^{i \gamma_5 \theta} \psi,\ \ \ \ \ \ \ \ \
  \overline{\psi} \rightarrow \overline{\psi} e^{i \gamma_5 \theta},
 $$
 $$
 \left( \matrix{\sigma \cr \pi \cr}\right) \rightarrow
 \left( \matrix{\hfill\cos 2\theta & \hfill\sin 2\theta\hfill\cr
-\sin 2\theta &\hfill\cos 2\theta \hfill\cr} \right)
 \left( \matrix{\sigma \cr \pi \cr}\right).\eqno(2.2)
$$
Using the action, we get the Langevin equations
\refmark\sakita
$$
\partial_t \psi_{\alpha}^i (x,t) =
    + \int d^4 y K_{\alpha \beta}(x,y)
     {\delta^r S \over \delta \overline{\psi}_{\beta}^i (y,t)}
    + \eta_{1\alpha}^i(x,t)
 + {1\over 2} \int d^4 y K_{\alpha \beta}(x,y) \eta_{2 \beta}^i(y,t)
             ,\eqno(2.3)
 $$
 $$
    \partial_t \overline{\psi}_{\alpha}^i (x,t)
      =
- \int d^4 y {\delta^r S \over \delta \psi_{\beta}^i (y,t)}
                                     K_{\beta \alpha}(y,x)
    + {1\over 2}
 \int d^4 y \overline{\eta}_{1 \beta}^i(y,t)K_{\beta \alpha}(y,x)
   + \overline{\eta}_{2 \alpha}^i (x,t),\eqno(2.4)
 $$
 $$
          K_{\alpha \beta}(x,y) \equiv
              \lbrace -i\gamma_{\mu}\partial_{\mu}
                   + g (\sigma  -i \gamma_5 \pi )
                   \rbrace _{\alpha \beta}\delta^4(x-y)
               ,
    $$
    $$
 \partial_t \pi (x,t) = - {\delta S \over \delta \pi (x,t)} + \zeta
(x,t),
   \eqno(2.5)
  $$
  $$
\partial_t \sigma (x,t) = - {\delta S \over \delta \sigma (x,t)}
+ \xi (x,t),\eqno(2.6)
$$
where $\alpha , \beta$ are spinor indices and
$K_{\alpha \beta}$ is a kernel which is introduced for the sake of
convergence of solutions in the equilibrium limit.
White noise fields $\eta,\overline{\eta},\zeta$ and $\xi$ are defined to
satisfy the following correlations
  $$
  \langle \eta_{\ell \alpha}^{i}(x,t)
       \overline{\eta}_{m \beta}^j(y,t') \rangle
   = 2 \delta^{ij} \delta_{\ell m} \delta_{\alpha \beta}
               \delta^4(x-y)\delta(t-t'),
  $$
  $$
        \langle \xi(x,t) \xi(y,t') \rangle =
      \langle \zeta(x,t) \zeta(y,t') \rangle =
            2\delta^4(x-y) \delta(t-t').
         $$
These Langevin equations and noise correlations are invariant under
eq.(2.2) and
 $$
 \eta_1 \rightarrow e^{i\gamma_5 \theta} \eta_1,\ \ \ \ \
 \eta_2 \rightarrow e^{-i\gamma_5 \theta} \eta_2,\ \ \ \ \
\overline{\eta}_1 \rightarrow \overline{\eta}_1\ e^{-i\gamma_5
\theta},
     \ \ \ \ \
\overline{\eta}_2 \rightarrow \overline{\eta}_2\ e^{i\gamma_5
\theta},
 $$
   $$
 \left( \matrix{\xi \cr \zeta \cr}\right) \rightarrow
  \left( \matrix{\hfill\cos 2\theta & \hfill\sin 2\theta\hfill\cr
       -\sin 2\theta &\hfill\cos 2\theta \hfill\cr} \right)
       \left( \matrix{\xi \cr \zeta \cr}\right).\eqno(2.7)
       $$
By using above Langevin equations, we obtain the integral equations
  $$
    \eqalign{\psi^i(x,t) = &\int^t_0 d\tau \ e^{\partial^2 (t-\tau)}
        \lbrack \eta_{1}^i(x,\tau)
  + {1 \over 2}\lbrace -i\gamma_{\mu} \partial_{\mu}
    + g (\sigma(\tau) -i\gamma_5 \pi(\tau)) \rbrace
\eta_{2}^i(x,\tau)
  \cr
    & + g\lbrace \gamma_5 \gamma_{\mu} \partial_{\mu}\pi
    + i\gamma_{\mu} \partial_{\mu}\sigma\rbrace \psi^i
   -g^2 \lbrace \sigma^2 + \pi^2 \rbrace \psi^i(\tau) \rbrack
      + e^{\partial^2 t}\psi^i(x,0),}\eqno(2.8)
    $$
     $$
              \eqalign{\overline{\psi}^i(x,t) =
            &
            \int^t_0 d\tau \ e^{ \partial^2(t-\tau)}
          \lbrack \overline{\eta}_{1}^i(x,\tau)
 {1 \over 2}\lbrace i\gamma_{\mu} \overleftarrow{\partial}_{\mu}
        + g(\sigma(\tau) -i\gamma_5 \pi(\tau))
           \rbrace
       +\overline{\eta}_2^i(x,\tau)
       \cr
      &
      + g \overline{\psi}^i(\tau)
      \lbrace
      \gamma_5 \gamma_{\mu} \partial_{\mu}\pi
           -
            i\gamma_{\mu} \partial_{\mu} \sigma
            \rbrace
    -g^2\lbrace \sigma^2
      + \pi^2\rbrace\overline{\psi}^i(\tau)\rbrack
      + e^{\partial^2 t} \overline{\psi}^i(x,0),}\eqno(2.9)
 $$
 $$
   \pi (x,t) = \int ^t_0 d\tau \ e^{ -N M^2 (t-\tau)}
     \lbrack \zeta (x,\tau)
     -
i g \overline{\psi}^i(x,\tau) \gamma_5 \psi^i (x,\tau)\rbrack
       + e^{-NM^2 t} \pi(x,0),\eqno(2.10)
      $$
      $$
    \sigma (x,t) = \int^t_0 d\tau \ e^{-NM^2(t-\tau)}
   \lbrack \xi (x,\tau)
   -
   g \overline{\psi}^i(x,\tau) \psi^i(x,\tau)\rbrack
     + e^{-NM^2 t} \sigma (x,0),\eqno(2.11)
 $$
where $\psi^i(x,0),\overline{\psi}^i(x,0),\pi(x,0)$ and $\sigma (x,0)$ are
initial values for the corresponding fields at {\it t = 0 }. These equations
can
be solved iteratively and solutions can be represented graphically as

$$
Fig.1
$$
where we denote $e^{\partial^2 (t-\tau)}$ in (2.8) and/or (2.9) by a solid
line, $e^{ -N M^2 (t-\tau)}$ in (2.10) by a wavy line, $e^{ -N M^2(t-\tau)}$
in (2.11) by a dotted line,
$\eta (\equiv \eta_1 -{i \over 2}\gamma_{\mu}\partial_{\mu}\eta_2)$
by a cross, $\eta_2$ by a encircled cross,
$\overline{\eta}(\equiv \overline{\eta}_1{i\over
2}\gamma_{\mu}\overleftarrow{\partial}_{\mu}+\overline{\eta}_2)$
by a box, $\overline{\eta}_1$ by a encircled box,
$\zeta$ by an asterisk and $\xi$ by a triangle.
In order to
take the large-N limit, we evaluate the degree of N in each stochastic
diagram. This evaluation enables us to get the ``effective Langevin
equation'' as follows.

 From above integral equations, it turns out that in a stochastic diagram,
each boson line (with or without the noise fields) carries the factor
${1 \over N M^2}$ and each fermion loop carries the factor $N$. So the degree
of N is
  $$
    degree \ of \ N = L_f - P_{\sigma} - P_{\pi},
    $$
in a stochastic diagram. Here $L_f$ is the number of fermion loops,
$P_{\sigma}$ is the number of $\sigma$ propagators and $P_{\pi}$ is the
number of $\pi$ propagators. Moreover from {\it Fig.}1, it turns out that there
is
$\eta$ or $\overline{\eta}$ in each edge of fermion line of vertices. For a
stochastic diagram with n external fermion lines, we obtain the following
relation
(see {\it Fig.}2).
  $$
    2 L_f + n \leq C_{\eta} + C_{\overline{\eta}}
     \leq 2 P_{\sigma} + 2 P_{\pi} +n,
    $$
    $$
    Fig.2
  $$
where $C_{\eta}(C_{\overline{\eta}})$ expresses the number of
$\eta_1(\overline{\eta}_1)$ and $\eta_2(\overline{\eta}_2)$
and in {\it Fig.}2 we denote $\langle \eta
\overline{\eta}\rangle$ by a box with cross. Therefore
  $$
    L_f - P_{\sigma} - P_{\pi} \leq 0.
    $$
So in the large-N limit, we should consider only the stochastic diagrams
which have the relation
  $$
    L_f - P_{\sigma} - P_{\pi} = 0.
   $$
Therefore in the large N limit, we should consider only the following
stochastic diagrams.
  $$
    Fig.3
  $$
Hereafter, we shall confine ourselves to the large-N limit. From {\it Fig.}3,
it
turns out that we may exchange the field $\sigma$ for its expectation value
$\sigma_c \equiv \langle \sigma \rangle $, which is independent of $x$, in the
Langevin equation. Likewise we may exchange the field $\pi$ for zero when we
choose $\langle \pi \rangle = 0$ at any fictitious time by means of the
symmetry
(2.2) and (2.7). After these exchanges, we obtain the ``effective Langevin
equations''
  $$
    \eqalign{\partial_t\psi^i(x,t) = (\partial^2 - m^2) & \psi^i(x,t) +
\eta_1^i
      + {1 \over 2} (- i \gamma_{\mu}\partial_{\mu} + m)
       \eta_2^i,
       \cr
    \partial_t\overline{\psi}^i(x,t) =
      (\partial^2 - m^2)& \overline{\psi}^i(x,t)
      +
      {1 \over 2} \overline{\eta}_1^i( i \gamma_{\mu}
      {\overleftarrow\partial}_{\mu} + m)
         + \overline{\eta}_2^i,
         \cr
      &
      m \equiv g \sigma_c,}\eqno(2.12)
  $$
and if $\partial_t\sigma_c(t)=0$, solutions of these equations
in momentum space are obtained as
  $$
    \eqalign{&\psi^i(k,t) = \int^t_0 d\tau \
        e^{ -(k^2 + m^2)(t-\tau) }
        \lbrack \eta_1^i(\tau)
         - {1 \over 2} (\gamma_{\mu} k_{\mu} - m) \eta_2^i(\tau) \rbrack
           + e^{ -(k^2 + m^2)t} \psi^i(k,0),
           \cr
    & \overline{\psi}^i(k,t) = \int^t_0 d\tau \
        e^{ -(k^2 + m^2)(t-\tau) }
        \lbrack
      {1 \over 2} \overline{\eta}_1^i(\tau) (\gamma_{\mu} k_{\mu} + m)
      + \overline{\eta}_2^i(\tau)  \rbrack
           + e^{ -(k^2 + m^2)t} \overline{\psi}^i(k,0).}\eqno(2.13)
  $$
We use these effective Langevin equations later, but we should comment in
justification of the assumption $\partial_t\sigma_c(t) = 0$ at first.

Because the fictitious time dependence of $\sigma$ is described by eq.(2.11),
we
calculate the expectation value of eq.(2.11)
  $$
     \sigma_c(t) = \int^t_0 d\tau \
      e^{ - NM^2(t-\tau)}
         (-g)\ \langle \overline{\psi}^i(\tau) \psi^i(\tau)\rangle
          + e^{ - NM^2 t} \sigma(0).\eqno(2.14)
       $$
We substitute eq.(2.13) into eq.(2.14), then we obtain
  $$
    \eqalign{\sigma_c(t) =
      -g \int^t_0 d\tau &\ e^{ - NM^2 (t-\tau)}
      \int {d^4k d^4k' \over (2 \pi)^8}
        e^{ -i(k + k')x}
        \lbrack -tr({- \gamma_{\mu}
         k_{\mu} + m \over k^2 + m^2}) N \delta(k+k')
         \cr
         &\times  \lbrace 1 - e^{ -2(k^2 + m^2 )\tau }
          \rbrace
          +
              e^{ -(k^2 + {k'}^2 +2m^2) \tau }
                \ \overline{\psi}^i(k,0) \psi^i(k',0) \ \rbrack
                \cr
         &\ \ \ \ \ \ \ \ \ \ \ \ \ \ + e^{ -NM^2 t } \sigma(0).}\eqno(2.15)
  $$
Performing the $\tau$ integration and taking account of large $NM^2$, this
equation becomes
  $$
    \eqalign{ \sigma_c (t)  =
      -{g \over NM^2} &\
      \int {d^4k d^4k' \over (2 \pi)^8}
        e^{ -i(k + k')x }
         \lbrack -tr({- \gamma_{\mu} k_{\mu} + m \over k^2 + m^2})N
         (2\pi)^4\delta^4(k+k') \cr
   &
         \times \lbrace 1 - e^{-2(k^2 + m^2 )t}
          \rbrace
            +  e^{ -(k^2 + {k'}^2 +2m^2) t }
                \ \overline{\psi}^i(k,0) \psi^i(k',0) \ \rbrack
    \cr
          + {g \over NM^2} &\ e^{ -NM^2t }
                \int {d^4k \ d^4k' \over (2\pi)^8}
              e^{-i(k+ k')x}
            \ \overline{\psi}^i(k,0) \psi^i(k',0)
       + e^{ -NM^2 t} \sigma(0),}
  $$
  $$
         \ \ \ \ \ \ \ \ \ \ \
         = {g \over M^2} \
      \int {d^4k \over (2 \pi)^4}
          tr({- \gamma_{\mu} k_{\mu} + m \over k^2 + m^2})
      +
       e^{ -NM^2t } \lbrace
      \sigma (x,0)
      +
      {g \over NM^2}\ \overline{\psi}^i\psi^i(x,0)
     \rbrace
     ,\eqno(2.16)
  $$
where we assume
$$
          \overline{\psi}^i(k,0)\psi^i(k',0)=
          -tr({-\gamma_{\mu}k_{\mu}+m \over k^2 + m^2})N(2\pi)^4\delta^4(k+k').
          \eqno(2.17)
$$
{}From (2.16), we obtain $\langle \partial_t\sigma(x,t)\rangle = 0$ if we
choose
the initial values as
 $\sigma(x,0)=-{g \over NM^2}\overline{\psi}^i\psi^i(x,0)$, i.e., we choose the
$\sigma_c$ as solution of the following equation
 $$
 \sigma_c(0)= {g \over M^2} \
      \int {d^4k \over (2 \pi)^4}
          tr({- \gamma_{\mu} k_{\mu} + g\sigma_c \over k^2 + g^2\sigma_c^2}).
          \eqno(2.18)
      $$
The contribution of the initial values is disappeared in the large-N limit, so
we
may choose them above. Therefore the assumption $\partial_t
\sigma_c = 0$ is consistent with the Langevin equations under the initial
condition (2.17) and (2.18).

For examples, using the effective Langevin equation (2.12) and (2.13) we
give the propagators of  $\sigma$ and $\pi$
  $$
  \langle \sigma'(k,t) \sigma'(k',t) \rangle_{t \rightarrow \infty}  =
    {\delta(k+k') \over NM^2} \lbrack 1
    - {4g^2 \over M^2}\int {d^4p \over (2\pi)^4}{(k-p)p
      + m^2 \over (p^2 + m^2) \lbrace (k-p)^2 + m^2 \rbrace}
       \rbrack ,\eqno(2.19)
    $$
      $$
        \sigma'(k,t) \equiv \sigma(k,t) - \sigma_c(t)
        $$
        $$
      \langle \pi(k,t) \pi(k',t) \rangle_{t \rightarrow \infty}
      =
        {\delta(k+k') \over NM^2}
       \lbrack 1
    +
       {4g^2 \over M^2}\int {d^4p \over (2\pi)^4}{-(k-p)p
   +
   m^2 \over (p^2 + m^2) \lbrace (k-p)^2 + m^2 \rbrace}
   \rbrack .\eqno(2.20)
  $$
The value of $m=g\sigma_c$ is determined from the expectation value of eq.(2.6)
  $$
    0 = \partial_t \sigma_c
    = -NM^2 \sigma_c
    -
    g\langle \overline{\psi}^i\psi^i\rangle
      =
         - NM^2 \sigma_c
      +
            gN\int {d^4k \over (2\pi)^4}
              tr\lbrack { -\gamma_{\mu} k_{\mu}+m \over k^2 +m^2 }
              \rbrack.\eqno(2.21)
       $$
If we choose M as our cutoff, we obtain the ``gap equation''
       $$
            M^2(1 - {g^2 \over 4 \pi^2})
     +
          {g^4 \over 4 \pi^2}\sigma_c^2 ln({M^2 \over g^2\sigma_c^2})
     =0. \eqno(2.22)
$$
Of course, these results conincide with ones which are obtained in
the path-integral method.

\chapter{Stability of vacuum}

Stability of vacuum can be discussed by eq.(2.21)
  $$
  \partial_t \sigma_c = -NM^2 \sigma_c
     -g\langle \overline{\psi}\psi\rangle. \eqno(3.1)
   $$
$\langle \overline{\psi}\psi\rangle$ can be evaluated by eq.(2.13) as
   $$
      \langle \overline{\psi}\psi\rangle =
   -
     {m^3 \over 4\pi^2} N \ \lbrack {M^2 \over m^2}
   -
        ln({M^2 \over m^2})\rbrack.\eqno(3.2)
         $$
Here we define the ``effective potential'' $U_{eff}$ as
  $$
    \partial_t\sigma_c(t)
    =
       - {\partial U_{eff}(\sigma_c) \over \partial \sigma_c},\eqno(3.3)
$$
and obtain the effective potential
  $$
     \eqalign{U_{eff}(\sigma_c,t) &= {1\over 2}NM^2 \sigma^2_c
        + g \int d\sigma_c\langle \overline{\psi}(x,t)\psi(x,t)\rangle \cr
    &=
        N\lbrack
           {1\over 2}M^2 (1-{g^2 \over 4\pi^2})\sigma_c^2
   +
        {1 \over 16\pi^2} g^4 \sigma_c^4
        \lbrace {1\over 2} + ln({M^2 \over g^2 \sigma_c^2})\rbrace
        \rbrack.}\eqno(3.4)
$$
where we omit an integrating constant.
This is the same one as ordinary effective potential. If $g^2 <
4\pi^2$, perturbative vacuum $\sigma_c = 0$ is stable. On the other hand,
if $g^2 > 4\pi^2$, perturbative vacuum $\sigma_c=0$ is unstable
and non-perturbative vacuum $\sigma_c \not= 0$, which is determined by
the gap equation, is stable (see {\it Fig.}4).
$$
   Fig.4
$$

Finally, we
comment about masses. Using $M^2 \gg 1$ and (2.20), inverse of $\pi$
propagator is obtained as

$$
      N \lbrack Z p^2 + M^2 (1-{g^2\over 4\pi})
      + {g^2 \over 4\pi^2}m^2 ln({M^2 \over m^2})\rbrack,\eqno(3.5)
$$
where {\it Z} is a renormalization constant
$$
   Z = {g^2 \over 8 \pi^2}\lbrace ln({M^2 \over m^2}) -1\rbrace
 $$
For $g^2>4\pi^2$, perturbative vacuum ($m=0$) is unstable because $\pi$
field becomes tachyon. On the other hand, non-perturbative vacuum
($m\not= 0$ determined from the gap equation) is stable and from the gap
equation, it turns out that $\pi$ field is massless. Likewise, for
$g^2>4\pi^2$ inverse of $\sigma$ propagator ($p^2=0$) is $Z (2m)^2$ and mass
of $\sigma$ field is $2m$. Moreover, from the effective Langevin equation it
turns out that {\it m} is the fermion mass.

\chapter{Summary}

  We give the method for non-perturbative study in SQM in this paper. We use
Langevin equations mainly and construct the effective Langevin equation for
Nambu$\cdot$Jona-Lasinio model. This effective Langevin equation is a
powerful tool for non-perturbative study and gives us the
effective potential. The effective potential coincides with the effective
potential
which is obtained in path-integral approach. Thus various results obtained in
path-integral method will be also obtained in SQM by means of the effective
Langevin equation or the effective potential.

If we can solve the Langevin equation (2.12) under another
assumption than $\partial_t\sigma_c=0$, we can construct the fictitious
time dependent effective potential and expect that symmmetry is restored at
small fictitious time and broken at large fictious time, i.e., symmetry
breaks down spontaneously as the fictitious time grows. In this case we
want to regard the fictitious time as the inverse of temperature, but this is a
further problem.

\chapter{Acknowledgments}
 We thank Dr.K.-I.Kondo and Dr.A.Nakamura for valuable discussion.

\vfill\eject

\refout

\vfill\eject

\chapter{Figure Captions}

Fig.1(a) Solution of (2.8)

Fig.1(b) Solution of (2.9)

Fig.1(c) Solution of (2.10)

Fig.1(d) Solution of (2.11)

Fig.2 Example of the stochastic diagram \hfill\break
$(n=2, L_f=1, P_{\sigma}=3, P_{\pi}=0,
C_{\eta}=C_{\overline{\eta}}=3)$

Fig.3 Remaining diagrams in the large-N limit

Fig.4 The effective potential ($g^2> 4\pi^2$)

\bye